\documentclass[
11pt,%
tightenlines,%
twoside,%
onecolumn,%
nofloats,%
nobibnotes,%
nofootinbib,%
superscriptaddress,%
noshowpacs,%
centertags]%
{revtex4}

\usepackage{ljm}
\usepackage{graphicx}
\usepackage{algorithm}
\usepackage{algorithmic}
\usepackage{cleveref}
\usepackage{tabularx}

\newcolumntype{Y}{>{\raggedleft\arraybackslash}X}

\begin{document}

\title{Partial Reuse AMG Setup Cost Amortization Strategy for the Solution of Non-Steady State Problems}
\author{\firstname{D.~E.}~\surname{Demidov}}
\email[E-mail: ]{dennis.demidov@gmail.com}
\affiliation{
    Kazan Branch of Joint Supercomputer Center,
    Scientific Research Institute of System Analysis, \\
    the Russian Academy of Sciences;
    2/31, Lobachevskii str., Kazan 420111 Russia}

%\firstcollaboration{(Submitted by A. M. Elizarov)} % Add if you know submitter.
%\lastcollaboration{ }

%\received{May 30, 2021; revised June 15, 2021; accepted June 25, 2021}

\begin{abstract}
    The partial reuse algebraic multigrid (AMG) setup cost amortization
    strategy is presented for the solution of non-steady state problems. The
    transfer operators are reused from the previous time steps, and the system
    matrices and the smoother operators are rebuilt on each of the AMG
    hierarchy levels. It is shown on the example of modelling a two-fluid dam
    break scenario that the strategy may decrease the AMG preconditioner setup
    cost by 40\% to 200\%. The total compute time is decreased by up to 20\%,
    but the specific outcome depends on the fraction of time that the setup
    step initially takes.
\end{abstract}


\maketitle

\section{Introduction}

Most of the numerical simulation problems today involve solution of large
sparse linear systems of equations obtained from discretization of partial
differential equations on either structured or unstructured meshes. The
combination of a Krylov subspace method with algebraic multigrid (AMG) as a
preconditioner is considered to be one the most effective choices for solution
of such systems~\cite{brandt1985algebraic,ruge1987algebraic,Trottenberg2001}.
However, the construction of an AMG preconditioner may take a considerable
fraction of the total compute time. Depending on the convergence rate of the
iterative solution, more than 50\% of the computations could be spent on the
setup. This cost is unavoidable when a steady state problem is being solved,
but it could be amortized for non-steady state problems either by reusing the
work that went into construction of the preconditioner.

The paper introduces the partial reuse AMG setup cost amortization strategy for
the solution of non-steady state problems. The AMG hierarchy is partially
updated on each of the time steps: the transfer operators (restriction and
interpolation) are kept from the previous time steps, while the system matrices
and the smoother operators on each level of the hierarchy are updated on every
iteration.  Since the computation of the transfer operators may take up to 50\%
of the setup, this allows to significantly reduce the setup time, while keeping
the preconditioner up to date with the latest system matrix changes. The
approach has been recently implemented in the opensource AMGCL
library~\cite{Demidov2019, Demidov2020}.

The partial reuse strategy is compared to the full reuse one, described
in~\cite{Demidov2012}. There the AMG preconditioner is used completely
unaltered for as many time steps as the solution is able to converge within a
reasonable number of iterations.  Reusing the preconditioner this way works
when the system matrix changes slowly over time, since there is a strong chance
that a preconditioner constructed for a specific time step will act as a
reasonably good preconditioner for a couple of subsequent time steps. However,
the convergence may deteriorate enough to neglect any time savings due to
reusing the preconditioner. It was shown in~\cite{Demidov2012} that this
strategy is mostly beneficial when the solution phase is comparable to or is
cheaper than the setup phase, which may be the case when the solution phase is
accelerated by using a GPU, or when just a couple of iterations are required
for the solution to converge.  Here it is shown on the example of solving a
two-fluid dam break scenario that the partial reuse strategy is able to
consistently reduce both the setup cost and the overall compute time, even when
the full reuse strategy is counterproductive.

The rest of the paper is structured as follows. \Cref{sec:amg} provides an
overview of the AMG method. \Cref{sec:strategies} describes the AMG setup cost
amortization strategies considered in the paper. \Cref{sec:experiments}
describes the numerical experiments used to test the efficiency of the
amortization strategies, and presents the results of the numerical experiments.

\section{Algebraic multigrid} \label{sec:amg}

This section describes the basic principles behind the
AMG~\cite{brandt1985algebraic, Stuben1999}.  The method solves a system of
linear algebraic equations \begin{equation} \label{eq:auf} Au = f,
\end{equation} where $A$ is a square matrix. Multigrid methods are based on
recursive application of the two-grid scheme, which combines \emph{relaxation}
and \emph{coarse grid correction}. Relaxation, or smoothing iteration $S$, is a
simple iterative method, such as damped Jacobi or Gauss--Seidel
iteration~\cite{barrett1994templates}. Coarse grid correction solves the
residual equation on the coarser grid, and improves the fine-grid approximation
with the interpolated coarse-grid solution. Transfer between the grids is
described with the \emph{transfer operators} $P$ (\emph{prolongation} or
\emph{interpolation}) and $R$ (\emph{restriction}).

\begin{algorithm}
    \caption{AMG setup.}
    \label{alg:setup}
    \begin{algorithmic}
        \STATE Start with the system matrix $A_1 = A$.
        \WHILE{the matrix $A_i$ is too large to be solved directly}
            \STATE Construct the transfer operators $P_i$ and $R_i$.
            \STATE Construct the smoother $S_i$.
            \STATE Construct the coarser system using Galerkin operator: $A_{i+1}
                = R_i A_i P_i$.
        \ENDWHILE
        \STATE Construct a direct solver for the coarsest system $A_L$.
    \end{algorithmic}
\end{algorithm}

In geometric multigrid methods the grid hierarchy, and the matrices $A_i$ and
operators $P_i$ and $R_i$ on each level of the hierarchy are supplied by the
user based on the problem geometry. In algebraic multigrid methods the grid
hierarchy and the transfer operators are in general constructed automatically,
based only on the algebraic properties of the matrix~$A$. Note that this step
may be both computationally intensive and hard to parallelize. \Cref{alg:setup}
describes the \emph{setup} phase of a generic AMG method. Here, the transfer
operators $P_i$ and $R_i$, as well as the smoother $S_i$ are constructed from
the system matrix $A_i$ on each level of the AMG grid hierarchy (a common
choice for the restriction operator $R$ is the transpose of the prolongation
operator $R=P^T$). The next coarser level of the AMG hierarchy is fully defined
by the transfer operators $P_i$ and $R_i$.  Usually the most time consuming
steps of the setup are the transfer operators construction and the evaluation
of the Galerkin operator.

\begin{algorithm}
    \caption{AMG V-cycle.}
    \label{alg:vcycle}
    \begin{algorithmic}
        \STATE Start at the finest level with an initial approximation
        $u_1 = u^0$.
        \WHILE{not converged}
            \FOR{each level of the hierarchy, finest-to-coarsest}
                \STATE Apply a couple of smoothing iterations to the current
                       solution: $u_i = S_i(f_i, u_i)$.
                \STATE Find residual $e_i = f_i - A_i u_i$ and
                       restrict it to the RHS on the coarser level:
                       $f_{i+1} = R_i e_i$.
            \ENDFOR
            \STATE Solve the coarsest system directly:
                   $u_L = A_L^{-1} f_L$.
            \FOR{each level of the hierarchy, coarsest-to-finest}
                \STATE Update the current solution with the interpolated
                       solution from the coarser level:
                       $u_i = u_i + P_i u_{i+1}$.
                \STATE Apply a couple of smoothing iterations to the current
                       solution: $u_i = S_i(f_i, u_i)$.
            \ENDFOR
        \ENDWHILE
    \end{algorithmic}
\end{algorithm}

After the AMG hierarchy has been constructed, it is used to solve the
system using a simple V-cycle shown in \Cref{alg:vcycle}. Usually AMG is not
used standalone, but as a preconditioner with an iterative Krylov subspace
method. In this case a single V-cycle is used as a preconditioning step.

\section{AMG setup cost amortization strategies} \label{sec:strategies}

This section describes possible AMG setup cost amortization strategies for the
solution of non-steady state problems. Generally speaking, a preconditioner for
the matrix $A$ should be an approximation for its inverse $A^{-1}$. However,
the approximation does not have to be perfect for the iterative Krylov solver
to work. During the solution of a non-steady state problem, the system matrices
at the adjacent time steps often have similar structure both in terms of the
non-zero pattern and the specific non-zero values of the matrix. Having this in
mind, it should be possible to reuse at least some of the work going into the
construction of an AMG preconditioner.

\begin{algorithm}
    \caption{Full reuse of the AMG hierarchy.}
    \label{alg:full}
    \begin{algorithmic}
        \FOR{each time step}
            \IF{first step \OR AMG hierarchy has to be rebuilt}
                \STATE Build AMG hierarchy using the current system matrix
            \ENDIF
            \STATE Solve the current system using the AMG preconditioner
        \ENDFOR
    \end{algorithmic}
\end{algorithm}

The full reuse strategy, introduced in~\cite{Demidov2012}, assumes that a
preconditioner constructed for a time step $k$ may be applicable for a number
of subsequent time steps. The decision on whether to reuse the current
preconditioner or construct a new one is based on the performance of the
preconditioner on the previous time step. When the solution on the last time
step was able to converge within the iteration limit specified by the user, the
preconditioner is deemed good enough and is reused. Otherwise, new
preconditioner is constructed for the current system matrix. The strategy is
outlined in~\Cref{alg:full}. It was shown in~\cite{Demidov2012} that it works
best when the solution step is accelerated with a GPU, which makes the relative
setup cost to increase significantly. Without the solution acceleration, the
full reuse strategy is often counterproductive, because any savings in the
setup cost are neglected by the increase in the solution time due to the
growing number of iterations.  Also, the outcome of the strategy highly depends
on the iteration threshold parameter specified by the user, which requires some
experimentation for each problem.

\begin{algorithm}
    \caption{Partial reuse of the AMG hierarchy.}
    \label{alg:partial}
    \begin{algorithmic}
        \FOR{each time step}
            \IF{first step}
                \STATE Build AMG hierarchy using the current system matrix
            \ELSE
                \STATE \emph{// Update the AMG hierarchy:}
                \FOR{all levels of the hierarchy}
                    \STATE Construct the smoother $S_i$ for the new system matrix
                           $A_i$.
                    \STATE Construct the coarser system using Galerkin operator:
                           $A_{i+1} = R_i A_i P_i$.
                \ENDFOR
                \STATE Construct a direct solver for the coarsest system $A_L$.
            \ENDIF
            \STATE Solve the current system using the AMG preconditioner
        \ENDFOR
    \end{algorithmic}
\end{algorithm}

As shown in~\cite{Demidov2012} and in the next section, the main assumption
behind the full reuse strategy does not always hold, and the system matrix may
change enough between the adjacent time steps for the unaltered preconditioner
reuse to be counterproductive. The partial reuse strategy proposed in this work
tries to reuse as much as possible from the existing AMG hierarchy as well as
to incorporate the data from the new system matrix. In order to achieve this,
the transfer operators on each level of the existing hierarchy are left
untouched, and the system matrices on all levels are updated with the Galerkin
operator $A_{i+1}=R_i A_i P_i$. This approach is especially well-suited for
non-smoothed aggregation AMG, since transfer operators here mostly depend on
the non-zero pattern of the system matrix, and not on its actual values. The
non-zero pattern of the matrices obtained with discretization of partial
differential equations on structured or unstructured grids rarely changes
between time steps. The partial reuse strategy is outlined
in~\Cref{alg:partial}.

The partial reuse strategy may still require a full rebuild occasionally, for
example if the matrix size changes between time steps as a result of adaptive
remeshing. Also, the quality of the updated preconditioner may deteriorate as
the system matrix diverges from its initial state, so it may be beneficial to
do preventive full rebuilds from time to time. This could be done either
regularly (every $m$-th time step) or using some convergence quality criteria,
similar to the full reuse strategy.

\section{Numerical experiments} \label{sec:experiments}

In order to test the cost amortization strategies described in the previous
section, the two-fluid dam break scenario was modelled using the Kratos
Multi-Physics framework~\cite{Dadvand2010,Dadvand2013}. The source code for the
example is available in~\cite{dambreak}. In the scenario, a water-filled cuboid
is positioned in one part of the domain. It is released at the start time and
the water spreads across the domain driven by gravity. The water hits an
obstacle and splashes develop. More details on the boundary conditions and
problem settings can be found in~\cite{Larese2008, Coppola2011}. There are
two basic substeps needed to evolve the solution from time step $k$ to time
step~$k+1$, each step requiring a solution of a linear system of equations:
\begin{enumerate}
    \item Find the motion in both phases as the solution of the two-fluid
        Navier--Stokes equations;
    \item Determine the position of the interface by solving a convection
        equation for the level-set function.
\end{enumerate}

The velocity and pressure fields of two incompressible fluids moving in the
domain $\Omega$ can be described by the incompressible two-fluid Navier--Stokes
equations
\begin{gather}
    \rho \left[ \frac{\partial \mathbf u}{\partial t} + (\mathbf u \cdot \nabla)
    \mathbf u \right] - \nabla \cdot \left[ 2 \mu \mathbf \varepsilon(\mathbf
    u) \right] + \nabla p = \mathbf f, \\
    \nabla \cdot \mathbf u = 0,
\end{gather}
where $\rho$ is the density, $\mathbf u$ is the velocity field, $\mu$ is the
dynamic viscosity, $p$ is the pressure, $\mathbf \varepsilon(\cdot)$ is the
symmetric gradient operator, and $\mathbf f$ is the external body force vector,
which includes the gravity and buoyancy forces, if required. The density,
velocity, dynamic viscosity, and pressure are defined as
\begin{equation}
    \mathbf u, p, \rho, \mu = \begin{cases}
        \mathbf u_1, p_1, \rho_1, \mu_1 \quad \mathbf x \in \Omega_1, \\
        \mathbf u_2, p_2, \rho_2, \mu_2 \quad \mathbf x \in \Omega_2,
    \end{cases}
\end{equation}
where $\Omega_1$ and $\Omega_2$ indicate the parts of $\Omega$ occupied by
fluids number 1 and 2 correspondingly.

Regarding the second step, the evolution of the fluid interface is updated
using the so-called level set method, which has been widely used to track free
surfaces in mould filling and other metal forming processes. The basic idea of
the level set method is to define a smooth scalar function $\psi(x, t)$, over
the computational domain $\Omega$ that determines the extent of subdomains
$\Omega_1$ and $\Omega_2$. For instance, positive values may be assigned to the
points belonging to $\Omega_1$, and negative values~--- to the points belonging
to $\Omega_2$. The position of the fluid front will be defined by the iso-value
contour $\psi(x, t) = 0$.  The evolution of the front $\psi = 0$ in any control
volume $V_t \subset \Omega$ that is moving with a divergence-free velocity
field $\mathbf u$ is described by the equation
\begin{equation}
    \frac{\partial \psi}{\partial t} + (\mathbf u \cdot \nabla) \psi = 0.
\end{equation}

Both the Navier--Stokes and the level set equations are discretized using a
stabilized finite element method with linear P1 elements for all the unknowns.
The resulting level set equation has 104,401 unknowns and 1,331,279 non-zero
elements in the system matrix. The Navier--Stokes equation has 417,604 unknowns
and 21,300,464 non-zero elements.  In total, 49 time steps are made in the
model example.  The level set equations and the Navier--Stokes equations across
the time steps are treated as independent non-steady state problems and are
used to test the efficiency of the AMG setup cost amortization strategies
considered here.

Both sets of problems are solved using the BiCGStab iterative
solver~\cite{barrett1994templates} preconditioned with non-smoothed aggregation
AMG. The damped Jacobi iteration is used as a smoother on each level of the AMG
hierarchy. Using the non-smoothed aggregation is beneficial in the case of the
partial reuse strategy, since the transfer operators in this case have minimal
dependence on the system matrix values, and mostly depend on the matrix
non-zero structure. The implementation uses the AMGCL
library~\cite{Demidov2019, Demidov2020}. The level set problem is solved using
the scalar (double precision) AMGCL backend, and the Navier--Stokes is solved
as a monolithic system using block-valued backend with $4 \times 4$ statically
sized matrices as the matrix values~\cite{Demidov2021}.  The solution in both
cases was parallelized with OpenMP and CUDA technologies. All tests were
conducted on a machine with 3.40GHz 4 core Intel Core i5-3570K CPU and NVIDIA
GeForce GTX 1050 Ti GPU.

\begin{table}
    \caption{Relative cost of the setup operations from \Cref{alg:setup}.}
    \label{tab:setup}
    \centering
    \begin{tabularx}{\textwidth}{l|YYYY}
        Step & Level set (OpenMP) & Level set (CUDA) & Navier--Stokes (OpenMP) &
        Navier--Stokes (CUDA) \\
        \hline
        Transfer operators $P_i$ and $R_i$        & 56\% & 23\% & 39\% & 15\% \\
        Galerkin operator $A_{i+1} = R_i A_i P_i$ & 36\% & 16\% & 44\% & 19\% \\
        Smoother $S_i$                            &  5\% &  3\% &  8\% &  6\% \\
        Direct solver for the coarsest system     &  2\% &  0\% &  0\% &  0\% \\
        Move hierarchy to the backend             &  0\% & 56\% &  0\% & 55\% \\
    \end{tabularx}
\end{table}

\Cref{tab:setup} shows the cost of each of the steps from \Cref{alg:setup} for
the level set and Navier--Stokes problems relative to the complete setup time.
The construction of the transfer operators takes significant portion of the
setup. The partial reuse strategy basically removes this step from the setup,
which allows to save 40 to 60\% of the setup cost.  When the CUDA backend is
used to accelerate the solution, an additional step of moving the constructed
hierarchy to the GPU memory appears in the setup, which consumes more than 50\%
of the setup time.  The relative cost of transfer operators construction grows
proportionally smaller in this case, which yields lower savings in the setup
time.

\begin{table}
    \caption{Cost savings for the full and partial reuse strategies.}
    \label{tab:strategies}
    \centering
    \begin{tabularx}{\textwidth}{l|YYYYYY}
        Strategy & Setup (s) & Solve (s) & Rebuilds
        & Average iterations & Total \newline speedup (\%)
        & Setup \newline speedup (\%) \\
        \hline
        & \multicolumn{6}{c}{Level set, OpenMP} \\
        No reuse      & 1.235 & 2.893 & 49 & 5.0 &      &         \\
        Full reuse    & 0.021 & 3.132 &  1 & 5.5 & 31\% & 5,781\% \\
        Partial reuse & 0.423 & 2.794 & 49 & 5.0 & 28\% &   192\% \\
        \hline
        & \multicolumn{6}{c}{Level set, CUDA} \\
        No reuse      & 2.064 & 0.944 & 49 & 5.0 &       &         \\
        Full reuse    & 0.037 & 0.904 &  1 & 5.5 & 220\% & 5,478\% \\
        Partial reuse & 0.949 & 0.775 & 49 & 5.0 &  75\% &   117\% \\
        \hline
        & \multicolumn{6}{c}{Navier--Stokes, OpenMP} \\
        No reuse      & 3.756 &  70.564 & 49 & 16.7 &  &  \\
        Full reuse    & 1.960 & 194.310 & 25 & 45.6 & -62\% & 92\% \\
        Partial reuse & 2.198 &  71.349 & 49 & 16.9 &   1\% & 71\% \\
        \hline
        & \multicolumn{6}{c}{Navier--Stokes, CUDA} \\
        No reuse      & 9.766 & 21.429 & 49 & 16.7 &  &  \\
        Full reuse    & 4.926 & 59.175 & 25 & 46.1 & -51\% & 98\% \\
        Partial reuse & 7.049 & 21.603 & 49 & 16.9 &   9\% & 39\% \\
    \end{tabularx}
\end{table}

\Cref{tab:strategies} compares the achieved cost savings for the no-reuse case
used as a baseline (the AMG is completely rebuilt from scratch on each of the
time steps), full reuse, and partial reuse strategies on the example of the
level set and the Navier--Stokes equations, solved with OpenMP and CUDA
backends. The columns show the setup and solve times in seconds (accumulated
across all of the time steps; the total compute time may be obtained as a sum
of these two columns), the number of full rebuilds of the AMG hierarchy, the
average number of iterations on a single time step, and the speedup observed
for the total compute time and the AMG setup time when using one of the
considered amortization strategies.

The results show that the full reuse strategy, as expected, may be a hit or
miss depending on the problem that is being solved.  Specifically, it appears
that the AMG preconditioner constructed for the level set problem at the first
time step is fully applicable to the rest of the time steps. The setup only had
to be performed once, which yields the total speedup of 31\% for the OpenMP
backend and 220\% for the CUDA-accelerated backend. However, for the
Navier--Stokes problem a preconditioner constructed for one time step shows bad
convergence on the next time step, which is obvious from the jump in the
average number of iterations (about 17 iterations in the no-reuse case vs 46
iterations in the full reuse case). In fact, the solution here was not able to
fully converge over the maximum number of 100 iterations on the every other
step, and the strategy has negative total speedup.

On the contrary, the partial reuse strategy shows consistently positive results
for both of the problems. The average number of iterations practically does not
grow, and the setup time is reduced by 39\% to 192\%. The total speedup in the
case of the level set problem is 28\% and 75\% for the OpenMP and CUDA backends
correspondingly. In the case of the Navier--Stokes problem, where the setup
takes only a small fraction of the total compute time, the observed total
speedup is 1\% and 9\%.

For both strategies, reusing the AMG setup makes more sense for the CUDA
backend, where the solution is about 3.5 times faster than on the OpenMP
backend. This makes the amortization strategy more important, since the
relative setup cost grows accordingly.

\section{Conclusion} \label{sec:conclusion}

The partial reuse AMG setup cost amortization strategy for the solution of
non-steady state problems introduced in this work is able to reduce the setup
cost of an AMG preconditioner by about 40\% to 200\%. It is shown on the
example of modelling the two-fluid dam break scenario that the strategy is able
to robustly decrease the total compute time, as opposed to the full reuse
strategy that may be counterproductive. The overall efficiency depends on the
relative cost of the setup with respect to the total compute time.  Naturally,
the approach outlined here makes more sense when the setup takes considerable
fraction of the compute time, which is usually the case when the solution step
is accelerated with a GPU, or the solution demonstrates fast convergence rate.

The AMG setup cost amortization strategy strategy proposed here allows to get
rid of one significant step in the AMG setup algorithm, namely the construction
of the transfer operators, which may be both expensive and serial in nature.
Another possibility to reduce the setup cost is to reuse the knowledge about
the system matrix non-zero pattern during the computation of the Galerkin
product $A_{i+1} = R_i A_i P_i$. This would get rid of the symbolic step in the
matrix-matrix product algorithm and potentially could halve the cost of the
partial setup. However, this remains outside the scope of the current work.

\section{Acknowledgements}

The work was carried out at the JSCC RAS as part of the government assignment.
The author would like to thank Dr. Riccardo Rossi for the helpful discussions.

%\bibliographystyle{spmpsci}
%bibliography{ref}


\begin{thebibliography}{10}
\providecommand{\url}[1]{{#1}}
\providecommand{\urlprefix}{URL: }
\expandafter\ifx\csname urlstyle\endcsname\relax
  \providecommand{\doi}[1]{DOI:~\discretionary{}{}{}#1}\else
  \providecommand{\doi}{DOI:~\discretionary{}{}{}\begingroup
  \urlstyle{rm}\Url}\fi

\bibitem{barrett1994templates}
R.~Barrett, M.~Berry, T. F. Chan, J.~Demmel, J.~Donato, J.~Dongarra,
  V.~Eijkhout, R.~Pozo, C.~Romine, and H.~Van~der Vorst, Templates for the solution
  of linear systems: building blocks for iterative methods.
\newblock SIAM,  1994.

\bibitem{brandt1985algebraic}
A.~Brandt, S.~McCormick, and J.~Huge, Algebraic multigrid ({AMG})
for sparse matrix
  equations.
\newblock Sparsity and its Applications \textbf{257} (1985).

\bibitem{Coppola2011}
H.~Coppola-Owen and  R.~Codina, A free surface finite element model
for low froude
  number mould filling problems on fixed meshes.
\newblock International J. for Numerical Methods in Fluids \textbf{66}(7),
  833--851 (2011).

\bibitem{Dadvand2013}
P.~Dadvand, R.~Rossi, M.~Gil, X.~Martorell, J.~Cotela, E.~Juanpere,
S. R.
  Idelsohn, and  E.~O{\~{n}}ate, {Migration of a generic multi-physics framework to
  HPC environments}.
\newblock Computers and Fluids \textbf{80}(1), 301--309 (2013).
\newblock \doi{10.1016/j.compfluid.2012.02.004}.

\bibitem{Dadvand2010}
P.~Dadvand, R.~Rossi, and  E.~O{\~{n}}ate, {An object-oriented
environment for
  developing finite element codes for multi-disciplinary applications}.
\newblock Archives of Computational Methods in Engineering \textbf{17}(3),
  253--297 (2010).
\newblock \doi{10.1007/s11831-010-9045-2}.

\bibitem{Demidov2019}
D.~Demidov, {AMGCL}: An efficient, flexible, and extensible algebraic multigrid
  implementation.
\newblock Lobachevskii J. of Mathematics \textbf{40}(5), 535--546 (2019).
\newblock \doi{10.1134/S1995080219050056}.

\bibitem{Demidov2020}
D.~Demidov, {AMGCL} -- a {C++} library for efficient solution of large sparse
  linear systems.
\newblock Software Impacts \textbf{6}, 100037 (2020).
\newblock \doi{10.1016/j.simpa.2020.100037}.

\bibitem{Demidov2021}
D.~Demidov, L.~Mu, and B.~Wang, Accelerating linear solvers for
{Stokes} problems
  with {C++} metaprogramming.
\newblock J. of Computational Science \textbf{49}, 101285 (2021).
\newblock \doi{10.1016/j.jocs.2020.101285}.
\newblock
  \urlprefix\url{https://www.sciencedirect.com/science/article/pii/S1877750320305809}.

\bibitem{Demidov2012}
D.~Demidov and  D.~Shevchenko, Modification of algebraic multigrid
for effective
  gpgpu-based solution of nonstationary hydrodynamics problems.
\newblock J. of Computational Science \textbf{3}(6), 460--462 (2012).

\bibitem{Larese2008}
A.~Larese, R.~Rossi, E.~O{\~n}ate, and S.~Idelsohn, Validation of
the particle
  finite element method ({PFEM}) for simulation of free surface flows.
\newblock Engineering Computations \textbf{25}(4), 385--425 (2008).

\bibitem{ruge1987algebraic}
J. W. Ruge and  K.~St{\"u}ben, Algebraic multigrid.
\newblock In: Multigrid methods, pp. 73--130. SIAM (1987).

\bibitem{Stuben1999}
K.~Stuben, {Algebraic multigrid (AMG): an introduction with applications}.
\newblock GMD Report~70, GMD, Sankt Augustin, Germany (1999).

\bibitem{Trottenberg2001}
U.~Trottenberg, C.~Oosterlee, and A.~Sch{\"{u}}ller, {Multigrid}.
\newblock Academic Press, London,  2001.

\bibitem{dambreak}
S.~von Wenczowski, Two-fluids dam break scenario.
\newblock
  \urlprefix\url{https://github.com/KratosMultiphysics/Examples/tree/master/fluid_dynamics/validation/two_fluid_dam_break}.
\newblock Accessed 15.05.2021.

\end{thebibliography}
\end{document}